\documentclass[runningheads]{svmult}

\usepackage{makeidx}   
\usepackage{graphicx}  
\usepackage{subeqnar}  
\usepackage{multicol}  
\usepackage{physmubb}  
\makeindex             
\usepackage{epsfig}
\def\GeV{{\rm GeV}}

\begin{document}
\title*{Global Fits of Parton Distributions}
\titlerunning{Global Fits of Parton Distributions}
%
\vspace{-1cm}
\author{Robert S. Thorne\\
Cavendish Laboratory, University of Cambridge, \\
Madingley Road, Cambridge, CB3 0HE, UK}

\authorrunning{Robert S. Thorne}

\maketitle              

\vspace{-1.2cm}

\section{Introduction}

The fundamental quantities one requires in the calculation of scattering 
processes involving hadronic particles are  the parton distributions. 
Global fits \cite{MRST2001}-\cite{ZEUSfit} use all available data, 
largely structure functions, and the most 
up-to-date QCD calculations, currently NLO--in--$\alpha_s(Q^2)$,  
to best determine these parton distributions and
their consequences. In the global fits input partons are parameterized as, e.g.
$$xf(x, Q_0^2) = (1-x)^{\eta}(1+\epsilon x^{0.5}+\gamma x)
x^{\delta}$$
at some low scale 
$Q_0^2 \sim 1-5 \GeV^2$, and evolved upwards using NLO
DGLAP equations. Perturbation theory should be valid if 
$Q^2 > 2 {\rm GeV^2}$, and hence one fits data
for scales above $2-5 \GeV^2$, and this cut should also remove the 
influence of 
higher twists, i.e. power-suppressed contributions. 

In principle there are many different parton distributions -- all quarks and
antiquarks and the gluons. However, $m_c, m_b \gg 
\Lambda_{{\rm QCD}}$ (and top does not usually contribute), 
so the heavy parton distributions are 
determined perturbatively. Also we assume $s=\bar s$, and that
isospin symmetry holds, i.e. $p \to n$ leads to $ d(x) \to u(x)$ and 
$u(x) \to d(x)$. This leaves 6 independent 
combinations. Relating $s$ to $1/2(\bar u + \bar d)$ we have the 
independent distributions
$$u_V = u- \bar u, \quad d_V =d-\bar d, 
\quad {\rm sea}=2*(\bar u + \bar d + 
\bar s), \quad \bar d - \bar u, \quad g. $$
It is also convenient to define 
$ \Sigma = u_V + d_V + {\rm sea} +(c+\bar c) +(b+\bar b)$.
There are then various sum rules constraining parton inputs and 
which are conserved by evolution order by order in 
$\alpha_S$, i.e. the number of up and down valence quarks   
and the momentum carried by partons (the latter being 
an important constraint on the gluon which is only probed indirectly), 
$$ \int_0^1 x\Sigma(x) +x g(x) \, dx =1. $$

When extracting partons one needs to consider that not only are there 
6 independent 
combinations, but there is also a wide distribution of
$x$ from $0.75$ to $0.00003$. 
One needs many different types of experiment for
a full determination.   
The sets of data usually used are: 
H1 and ZEUS $F^p_2(x,Q^2)$ data \cite{H1A,ZEUS} which covers 
small $x$ and a wide range of $Q^2$; E665 
$F^{p,d}_2(x,Q^2)$ data \cite{E665} at medium $x$;
BCDMS and SLAC $F^{p,d}_2(x,Q^2)$ data \cite{BCDMS}-\cite{SLAC} at large $x$; 
NMC $F^{p,d}_2(x,Q^2)$ \cite{NMC} at medium and large $x$; 
CCFR $F^{\nu(\bar\nu) p}_2(x,Q^2)$ and
$F^{\nu(\bar\nu) p}_3(x,Q^2)$ data \cite{CCFR} at large $x$ 
which probe the singlet and valence quarks independently;
ZEUS  and H1 $F^{p}_{2,charm}(x,Q^2)$ data \cite{ZEUSc,H1c}; 
E605 $ p N \to \mu \bar \mu + X$ \cite{E605} constraining the large $x$ 
sea; E866 Drell-Yan asymmetry \cite{E866} which determines  
$\bar d -\bar u$; CDF W-asymmetry data \cite{Wasymm} 
which constrains the $u/d$ ratio at 
large $x$; CDF and D0 inclusive jet data \cite{D0,CDF} 
which tie down the high $x$ gluon; 
and NuTev Dimuon data \cite{NuTeV}
which constrain the strange sea.

The quality of the fit to data is usually determined by the $\chi^2$. 
There are various alternatives for calculating this.
The simplest is adding statistical and systematic errors in quadrature. 
This ignores the correlations between data points, but it is the only 
available method for many data sets. In principle it should be improved
upon, but in practice sometimes works perfectly well. 

A more sophisticated approach is to use the covariance matrix
$$C_{ij} = \delta_{ij} \sigma_{i,stat}^2 + \sum_{k=1}^n \rho^k_{ij}
\sigma_{k,i}\sigma_{k,j},$$
where $k$ runs over each source of correlated systematic error
and $\rho^k_{ij}$ are the correlation coefficients. The $\chi^2$ 
is defined by
$$\chi^2 = \sum_{i=1}^N\sum_{j=1}^N (D_i-T_i(a))C^{-1}_{ij}
(D_j-T_j(a)),$$
where $N$ is the number of data points, $D_i$ is the 
measurement
and $T_i(a)$ is the theoretical prediction depending on parton input 
parameters $a$. Unfortunately, this relies on inverting large matrices. 

One can also minimize with respect to the 
systematic errors, i.e. incorporate 
the systematic errors into the theory prediction
$$ f_i(a,s) = T_i(a) + \sum_{k=1}^n s_k \Delta_{ik},$$
where $\Delta_{ik}$ is the one-sigma correlated error for point 
$i$ from source $k$. In this case the $\chi^2$ is defined by     
$$\chi^2 = \sum_{i=1}^N \biggl(\frac{D_i-f_i(a,s)}{\sigma_{i,unc}}
\biggr)^2 + \sum_{k=1}^n s_k^2,$$
where the second term constrains the values of $s_k$. This allows 
the data to move {\it en masse} relative to the theory, but assumes the 
correlated systematic errors are Gaussian distributed.
One can actually solve for each of the $s_k$ analytically \cite{CTEQLag}, 
simplifying greatly.
This method is identical to the correlation matrix definition 
of $\chi^2$ at the minimum, 
but it has the double advantage that smaller matrices need 
inverting and one sees explicitly the shift of data relative to theory.  
However, one may ask whether Gaussian correlated errors are realistic and
whether it is valid to move data to compensate for the shortcomings of theory. 
MRST find that for HERA data increments in $\chi^2$ 
using this method are much the same as for adding errors in quadrature, 
and data move towards theory \cite{MRST2001}.
However, for Tevatron jet data the correlated systematic errors 
dominate and must be incorporated properly. 

Once a decision about $\chi^2$ is made, the above procedure completely 
determines parton distributions at present.  
The total fit is reasonably good and that for CTEQ6 \cite{CTEQ6} is shown in 
Table 1 for the large data sets. 
The total $\chi^2=1954/1811$. For MRST  
The total $\chi^2=2328/2097$
-- but the errors are treated differently, and different data sets and cuts
are used. The same sort of conclusion is true for other 
{\it global fits} \cite{Botje,Giele,Alekhin,H1Krakow,ZEUSfit}  
(which use fewer data). 
However, there are some areas where the theory perhaps needs to be improved, 
as we will discuss later.

\vspace{-0.5cm}

\begin{table}
\caption{Quality of fit to data for CTEQ6M}
\begin{center}
\renewcommand{\arraystretch}{1.4}
\setlength\tabcolsep{5pt}
\begin{tabular}{lll}
\hline\noalign{\smallskip}
Data Set & no. of data & $\chi^2$\\ 
\noalign{\smallskip}
\hline
\noalign{\smallskip}
H1 $ep$           & 230      & 228   \\                   
ZEUS $ep$         & 229      & 263    \\                   
BCDMS $\mu p$     & 339      & 378    \\
BCDMS $\mu d$     & 251      & 280    \\                   
NMC $\mu p$       & 201      & 305   \\                                
E605 (Drell-Yan) & 119      &  95   \\        
D0 Jets           &  90      &  65  \\ 
CDF Jets          &  33      &  49  \\
\hline
\end{tabular}
\end{center}
\label{Tab1a}
\end{table}

\vspace{-1.3cm}

\section{Parton Uncertainties} 

\vspace{-0.2cm}

\subsection{Hessian (Error Matrix) approach} 

In this one defines the Hessian matrix $H$ by
$$ \chi^2 -\chi_{min}^2 \equiv \Delta \chi^2 = \sum_{i,j} 
H_{ij}(a_i -a_i^{(0)})
(a_j -a_j^{(0)}). $$
$H$ is related to the covariance matrix 
of the parameters by
$C_{ij}(a) = \Delta \chi^2 (H^{-1})_{ij},$
and one can use the standard formula for linear error propagation.  
$$(\Delta F)^2 = \Delta \chi^2 \sum_{i,j} \frac{\partial F}
{\partial a_i}(H)^{-1}_{ij}  
\frac{\partial F}{\partial a_j},$$
This has been employed to 
find partons with errors by Alekhin \cite{Alekhin}
and H1 \cite{H1Krakow} (each with restricted data sets), as demonstrated in 
Fig.~1. 

\begin{figure}
\begin{center}
\includegraphics[width=.6\textwidth]{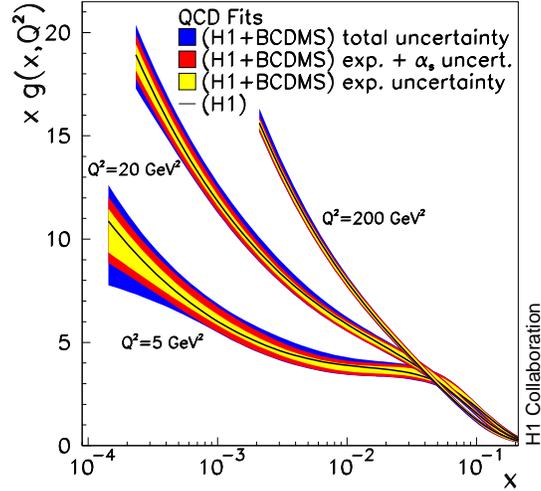}
\end{center}
\vspace{-0.4cm}
\caption[]{H1 determination of the gluon from their own data + BCDMS data
with an emphasis on $g(x,Q^2)$ and $\alpha_S(M_Z^2)$ in the fit}
\label{eps2}
\end{figure}

\vspace{-0cm}

\begin{figure}
\begin{center}
\includegraphics[width=.55\textwidth]{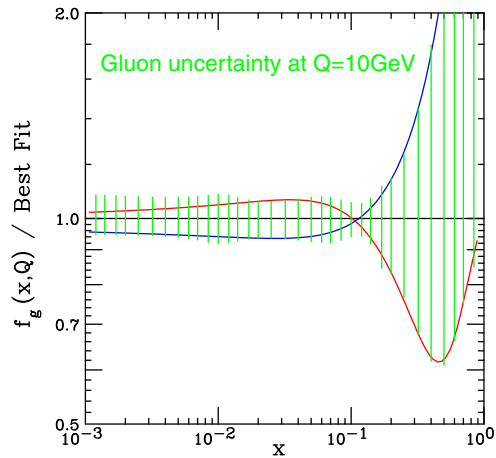}
\end{center}
\vspace{-0.4cm}
\caption[]{Results of CTEQ Hessian approach for the gluon uncertainty}
\vspace{-0.5cm}
\label{eps8}
\end{figure}

The simple method can be problematic with larger data sets and 
numbers of parameters due to extreme 
variations in $\Delta \chi^2$ in different directions in parameter 
space. This is solved by finding and rescaling the eigenvectors of $H$ 
(CTEQ \cite{CTEQmul,CTEQHes,CTEQ6}) leading to the diagonal form 
$$\Delta \chi^2 = \sum_{i} z_i^2. $$
The uncertainty on a physical quantity is then given by
$$(\Delta F)^2 = \sum_{i} \bigl(F(S_i^{(+)})-F(S_i^{(-)})\bigr)^2,$$
where $S_i^{(+)}$ and $S_i^{(-)}$ are PDF sets 
displaced along eigenvector
directions by a given $\Delta \chi^2$. Similar eigenvector 
parton sets have also been introduced by MRST \cite{MRSTnew}.  
However, there is an art in choosing the ``correct'' 
$\Delta \chi^2$ given the complication of the errors in the full 
fit \cite{THJPG}. 
Ideally $\Delta \chi^2 = 1$, but this leads to unrealistic errors. 
CTEQ choose $\Delta \chi^2 
\sim 100$, which is perhaps conservative. 
MRST choose $\Delta \chi^2 \sim 50$. An example of results 
is shown in Fig.~2.

\vspace{-1.9cm}

\begin{figure}
\begin{center}
\includegraphics[width=.7\textwidth]{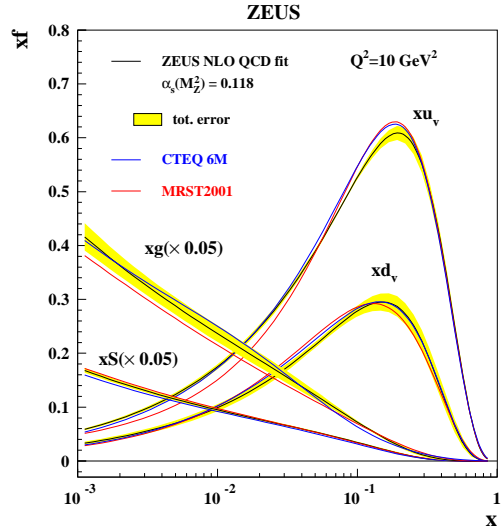}
\end{center}
\vspace{-2.4cm}
\caption[]{Parton densities and their errors extracted by fits by ZEUS}
\label{eps9}
\end{figure}

\vspace{-1cm}

\subsection{Offset method}  

In this the best fit is obtained by minimizing
$$\chi^2 = \sum_{i=1}^N \biggl(\frac{(D_i-T_i(a))}{\sigma_{i,unc}}
\biggr)^2,$$
i.e. the best fit and parameters $a_0$ are obtained using only uncorrelated 
errors, forcing the theory to be close to 
unshifted data. The quality of the fit is  
estimated by adding errors in quadrature. The systematic errors on the  
$a_i$ are determined by letting each $s_k = \pm 1$ and adding the  
deviation in quadrature, or equivalently by calculating 2 Hessian matrices
$$ M_{ij} = \frac {\partial ^2 \chi^2}{\partial a_i \partial a_j}
\qquad  V_{ij} = \frac {\partial ^2 \chi^2}{\partial a_i \partial s_j},$$
and defining covariance matrices  
$$C_{stat} = M^{-1} \qquad C_{sys} = M^{-1}VV^TM^{-1} \qquad C_{tot} 
= C_{stat} + C_{sys},$$
which is used in practice. This was used in early H1 \cite{Zomer} 
and ZEUS \cite{Botje} fits.  
It is still used by ZEUS \cite{ZEUSfit}, as shown in  Fig.~3, and is a 
conservative approach to 
systematic errors leading to a bigger uncertainty for a given $\Delta \chi^2$.

\vspace{-0.2cm}

\subsection{Statistical Approach} 

In principle one 
constructs an ensemble of distributions labelled by ${\cal F}$ each with 
probability $P(\{{\cal F}\})$, where one can incorporate the full 
information about measurements and their error 
correlations into the calculation of $P(\{{\cal F}\})$. 
This is statistically correct, and does not rely on 
the approximation of linear propagation errors in calculating observables.
However, it is inefficient, and in practice 
one generates $N_{pdf}$ ($N_{pdf}$ can be as low as $100$) 
different distributions
with unit weight but distributed according to $P(\{{\cal F}\})$ \cite{Giele}. 
Then the mean $\mu_O$ and deviation $\sigma_O$ of an observable $O$ are  
given by
$$\mu_O = \frac {1}{N_{pdf}}\sum_1^{N_{pdf}}O(\{{\cal F}\}), \quad
\sigma_O^2 =\frac {1}{N_{pdf}} \sum_{1}^{N_{pdf}}(O(\{{\cal F}\})-\mu_O)^2.
$$

Currently this approach uses only proton DIS data sets in order to avoid 
complicated 
uncertainty issues, e.g. shadowing effects for nuclear targets, 
and also demands consistency between data sets.
However, it is difficult to find 
many truly compatible DIS experiments, and consequently the 
Fermi2001 partons are determined by only 
H1, BCDMS, and E665 data sets. 
They result in good predictions for many Tevatron cross-sections. 
However, the restricted data sets mean there is restricted information --
data sets are deemed either perfect or, in the case of  most of them,  
useless -- leading to unusual values for some parameters. e.g.  
$\alpha_S(M_Z^2)=0.112 \pm 0.001$ and a very hard $d_V(x)$ 
at high $x$ (together these facilitate a good fit to Tevatron jets 
independent of the high-$x$ gluon). 
These partons would produce some extreme predictions.
Nevertheless, the approach does demonstrate that the Gaussian approximation 
is often not good, and therefore highlights shortcomings in the methods 
outlined in the previous sections.
It is a very attractive, but ambitious large-scale project, still in need 
of some further development, in particular the inclusion of 
a wider variety of data.    

\vspace{-0.2cm}

\subsection{Lagrange Multiplier method} 

This was first suggested by CTEQ \cite{CTEQLag} and 
has been concentrated on by MRST \cite{MRSTnew}. One performs the 
fit while constraining the value of some physical quantity, i.e. one minimizes 
$$ \Psi(\lambda,a) = \chi^2_{global}(a)  + \lambda F(a)$$
for various values of $\lambda$. This gives a set of best fits for 
particular 
values of the quantity $F(a)$ without relying on the quadratic approximation 
for $\chi^2$. 
The uncertainty is then determined by deciding an allowed range of 
$\Delta \chi^2$.
One can also easily check the variation in $\chi^2$ 
for each of the experiments in the global fit and ascertain if the total
$\Delta\chi^2$ is coming specifically from one region, which might cause 
concern. In principle, this is superior to the Hessian approach, but it must 
be repeated for each physical process. 

\vspace{-0.2cm}

\subsection{Results}

I choose the cross-section for $W$ and Higgs production at the Tevatron and 
LHC (for $M_H=115\GeV$) as examples.  
Using their fixed value of $\alpha_S(M_Z^2) =0.118$ and $\Delta \chi^2=100$ 
CTEQ obtain 
$$\Delta \sigma_{W}(\rm LHC) \approx \pm 4\% \quad 
\Delta \sigma_{W}({\rm Tev}) \approx \pm 5\% \quad \Delta 
\sigma_{H}({\rm LHC}) \approx \pm 5\%.$$
Using a slightly wider range of data, 
$\Delta \chi^2 \sim 50$ and $\alpha_S(M_Z^2) =0.119$ MRST obtain
$$\Delta \sigma_{W}(\rm Tev) \approx \pm 1.2\% \quad 
\Delta \sigma_{W}({\rm LHC}) \approx \pm 2\%$$
$$\Delta \sigma_{H}(\rm Tev) \approx \pm 4\% \quad 
\Delta \sigma_{H}({\rm LHC}) \approx \pm 2\%.$$
MRST also allow $\alpha_S(M_Z^2)$ to be free. 
In this case $\Delta \sigma_{W}$ is quite 
stable but $\Delta \sigma_{H}$ almost doubles. Contours of variation in 
$\chi^2$ for predictions of these cross-sections are shown in Fig.~4.    

\vspace{-0.3cm}

\begin{figure}
\includegraphics[width=.43\textwidth]{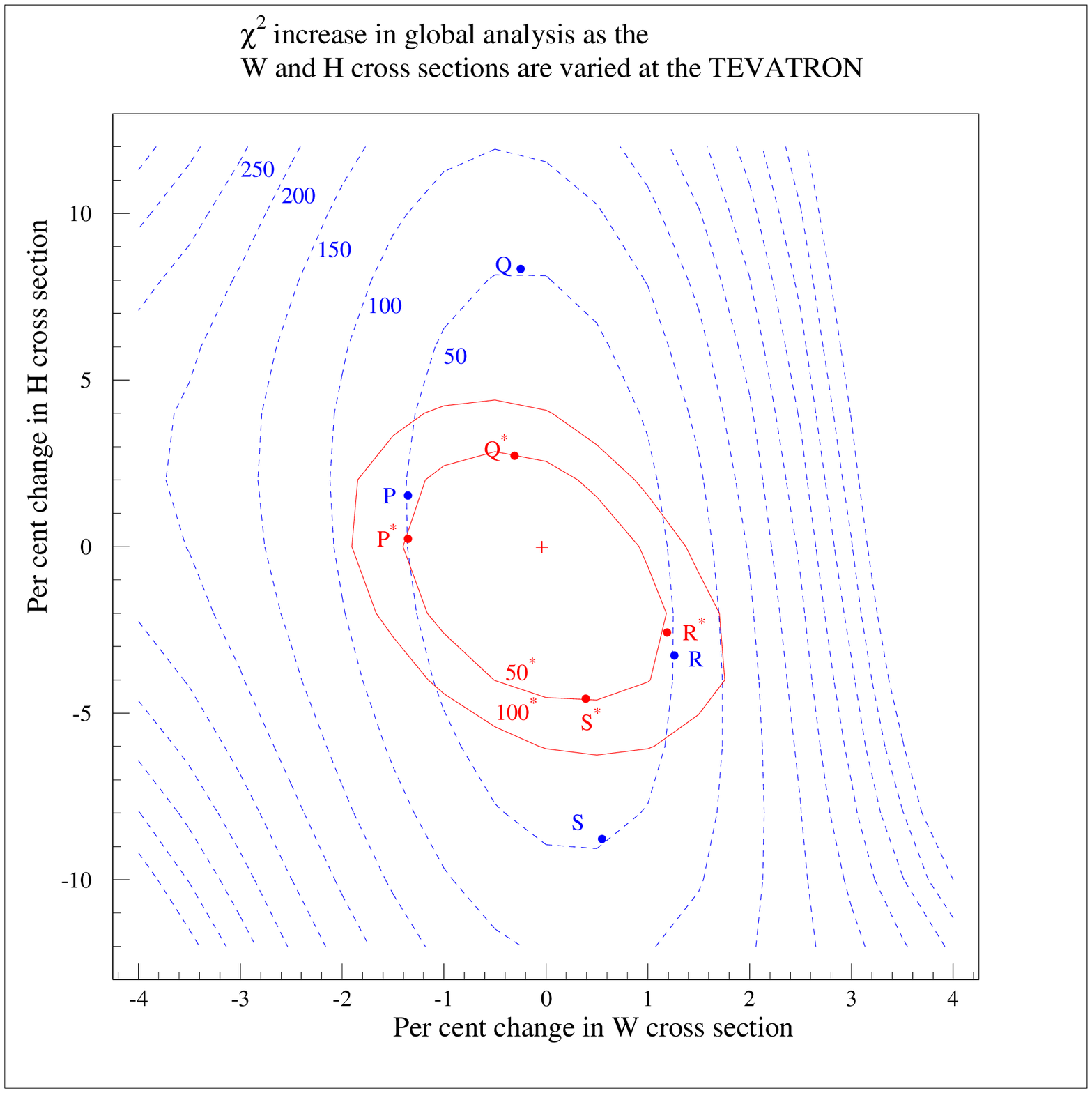}
\hspace{1cm}
\includegraphics[width=.43\textwidth]{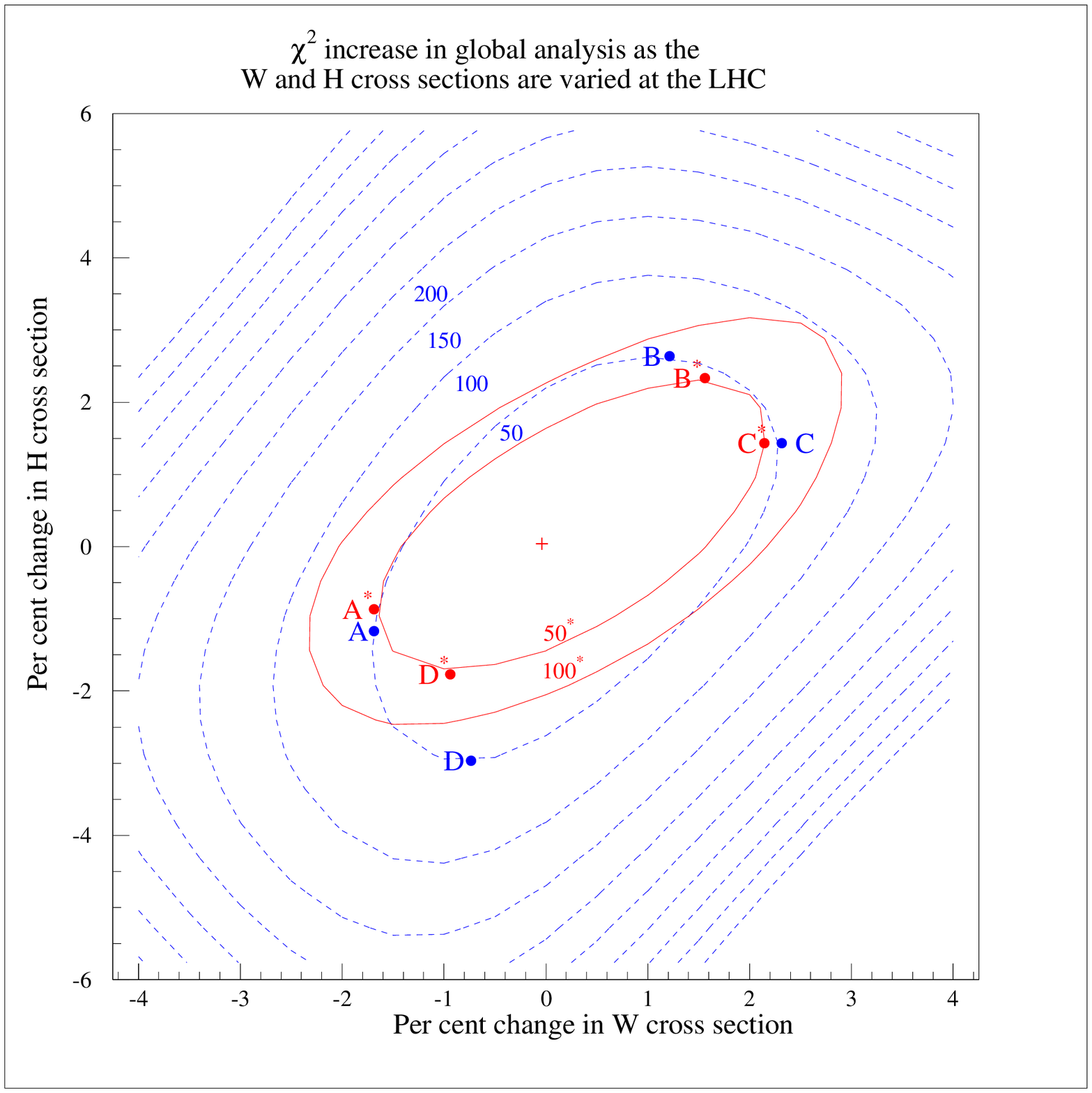}
\vspace{-1.4cm}
\caption[]{$\chi^2$-plot for $W$ and Higgs production at the 
Tevatron (left) and LHC (right) 
with $\alpha_S$ free and fixed at $\alpha_S=0.119$}
\label{eps15}
\end{figure}

\vspace{-0.4cm}

Hence, the estimation of uncertainties due to experimental errors 
has many different approaches and different types and amount of data actually 
fit. Overall the uncertainty from this source 
is rather small --  
only more than a few $\%$ for quantities determined by  
the high $x$ gluon and very high $x$ down quark.
However, different approaches can lead to rather different central values,
as illustrated for determinations of $\alpha_S(M_Z^2)$ in Table 2. This shows
that there are other matters to consider.   
As well as the experimental errors on data we need to determine the effect of 
assumptions made about the fit, e.g. 
cuts made on the data, the data sets fit, the parameterization for input sets, 
the form of the strange sea, {\it etc}.. 
Many of these can be as important as the 
errors on the data used (or more so). This is demonstrated in Fig.~5
which shows the predictions for 
$W$ and Higgs production at the Tevatron from MRST2001 and CTEQ6. As well as 
the consequences of these assumptions we must also consider 
the related problem of theoretical errors.

\vspace{-0.5cm}

\begin{table}
\caption{Values of $\alpha_s(M_Z^2)$ and its error from different NLO QCD 
fits}
\begin{center}
\renewcommand{\arraystretch}{1.4}
\setlength\tabcolsep{5pt}
\begin{tabular}{lll}
\hline\noalign{\smallskip}
Group & $\Delta \chi^2$ & $\alpha_S(M_Z^2)$ \\ 
\noalign{\smallskip}
\hline
\noalign{\smallskip}
CTEQ6 & $\Delta \chi^2 = 100$ & $0.1165
  \pm 0.0065(exp) $ \\
  ZEUS  & $\Delta \chi_{eff}^2 = 50$ &$ 0.1166
  \pm 0.0049(exp)\pm 0.0018(model)$ $\pm 0.004(theory) $ \\
  MRST01 & $\Delta \chi^2 = 20$ 
  &$ 0.1190 \pm 0.002(exp)\pm 0.003(theory)  $ \\
  H1    & $\Delta \chi^2 = 1$ &  
  $0.115 \pm 0.0017(exp) ^{+~~0.0009}_{-~~0.0005}~(model)$ 
  $\pm 0.005(theory)$ \\
  Alekhin & $\Delta \chi^2 = 1$ & $0.1171 
  \pm 0.0015(exp) \pm 0.0033(theory)$ \\
 GKK & $\Delta \chi_{eff}^2 = 1$  & $0.112 \pm 0.001(exp)$ \\

\hline
\vspace{-0.5cm}
\end{tabular}
\end{center}
\label{Tab1b}
\end{table}

\vspace{-1.4cm}

\begin{figure}
\begin{center}
\includegraphics[width=.5\textwidth]{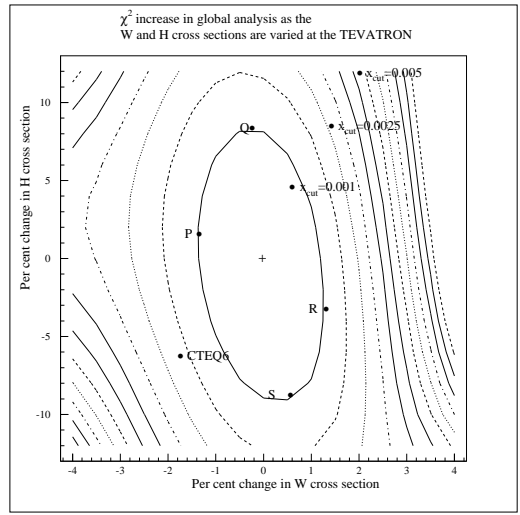}
\end{center}
\vspace{-1.9cm}
\caption[]{$\chi^2$-plot for $W$ and Higgs production at the Tevatron
with $\alpha_S$ free. The predictions from CTEQ6 and for fits with only data 
with $x>x_{cut}$ retained are marked}
\label{eps34}
\end{figure}

\vspace{-1.1cm} 

\section{Theoretical errors}

\vspace{-0.2cm}

\subsection{Problems in the fit}

\begin{figure}
\begin{center}
\includegraphics[width=.49\textwidth]{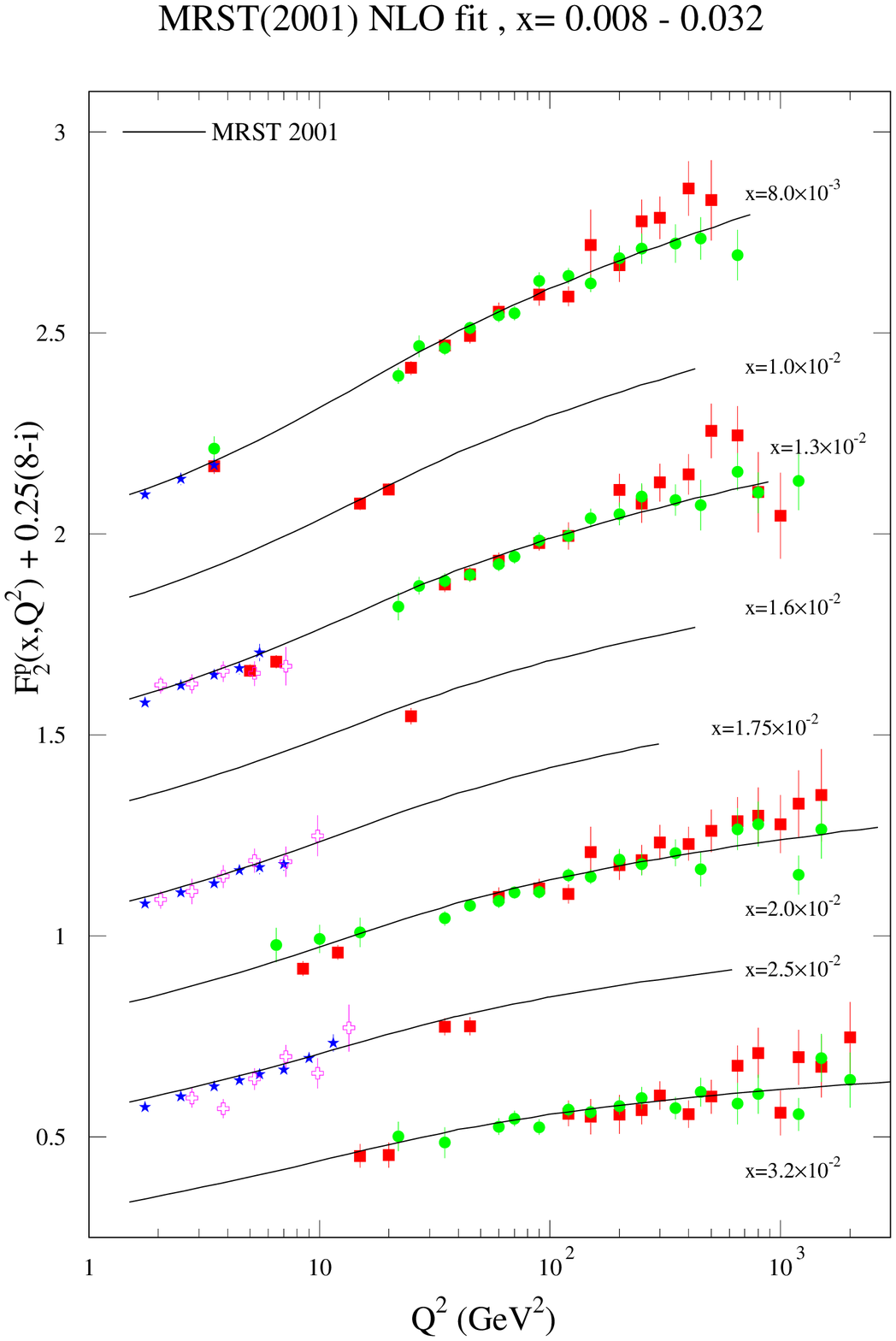}
\includegraphics[width=.49\textwidth]{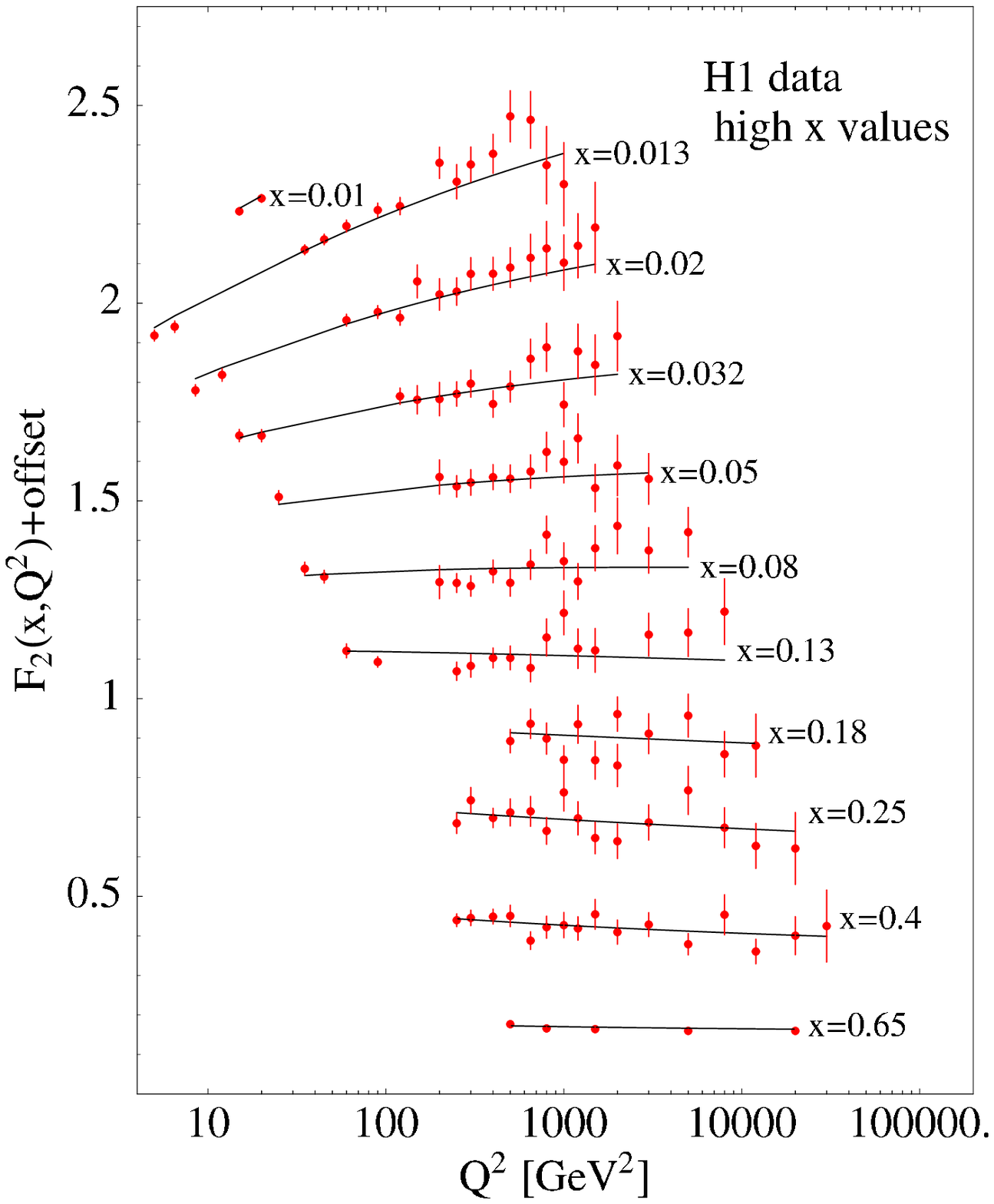}
\end{center}
\vspace{-0.3cm}
\caption[]{Comparison of MRST(2001) $F_2(x,Q^2)$ with HERA,
NMC and E665 data (left) and  CTEQ6 $F_2(x,Q^2)$ with H1 data
right}
\vspace{-0.3cm}
\label{eps21}
\end{figure}
 
\vspace{-0.1cm}

Theoretical errors are indicated by some regions where the theory perhaps 
needs to be improved to fit the data better.
There is a reasonably good fit to HERA data, 
but there are some problems at the highest 
$Q^2$ at moderate $x$, i.e. in $d F_2/d \ln Q^2$, as seen for MRST and CTEQ 
in Fig.~6. Also the data require the gluon to be valencelike or negative at 
small $x$ at low $Q^2$, e.g. the ZEUS gluon in Fig.~7, leading to 
$F_L(x,Q^2)$ being negative at the smallest $x,Q^2$ \cite{MRST2001}.
However, it is not just the low $x$--low $Q^2$ data that require 
this negative gluon. 
The moderate $x$ data need lots of gluon to get a reasonable 
$d F_2/d \ln Q^2$ and the Tevatron jets need a large high $x$ gluon, 
and this must be compensated for elsewhere.  
In general MRST find that it is difficult to reconcile the fit to 
jets and to the rest of the data, and that   
different data compete over the gluon and $\alpha_S(M_Z^2)$.
The jet fit is better for CTEQ6 largely due to their 
different cuts on other data. Other fits do not include the Tevatron jets,
but generally produce gluons largely incompatible with this data.

\vspace{-0.3cm}

\subsection{Types of Theoretical Error, NNLO}

It is vital to consider theoretical errors. These include 
higher orders (NNLO), small $x$ ($\alpha_s^n \ln^{n-1}(1/x)$),
large $x$ ($\alpha_s^n \ln^{2n-1}(1-x)$)
low $Q^2$ (higher twist), {\it etc}..
Note that renormalization/factorization 
scale variation is not a reliable  method of estimating these 
theoretical errors because of increasing 
logs at higher orders, e.g. at small $x$
$$P^1_{qg} \sim \alpha_S(\mu^2) \qquad \qquad P^2_{qg} 
\sim \frac{\alpha_s(\mu^2)}{x} \qquad P^n_{qg} 
\sim \frac{\alpha_S^n(\mu^2) \ln^{n-2}(1/x)}{x}$$
and scale variations of $P^1_{qg}, P^2_{qg}$ 
never give an indication of these logs.

\begin{figure}
\begin{center}
\includegraphics[width=.6\textwidth]{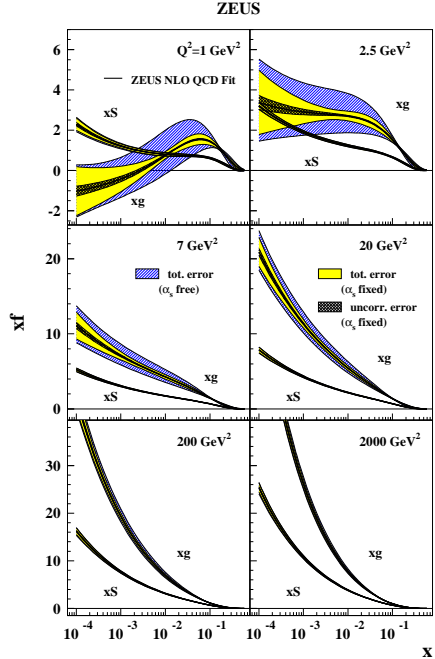}
\end{center}
\vspace{-0.72cm}
\caption[]{Zeus gluon and sea quark distributions at various $Q^2$ values}
\label{eps24}
\vspace{-0.4cm}
\end{figure}

\vspace{-0cm}

In order to investigate the true theoretical error we 
must consider some way of 
performing correct large and small $x$ resummations,
and/or use what we already know about NNLO. 
The coefficient functions are known at NNLO.
Singular limits $x \to 1$, $x \to 0$ are known for NNLO
splitting functions as well as limited moments \cite{NNLOmoms}, and this
has allowed approximate NNLO 
splitting functions to be devised \cite{NNLOsplit}
which have been used in approximate global fits \cite{MRSTNNLO}.
They improve the quality of fit very slightly  
(mainly at high $x$) and 
$\alpha_S(M_Z^2)$ lowers from $0.119$ to 0.1155. 
The gluon is smaller at NNLO at low $x$ due to the positive NNLO 
quark-gluon splitting function. 
There is also a NNLO fit by Alekhin \cite{NNLOAl}, with some differences 
-- the gluon is not smaller, probably due 
to the absence of Tevatron jet data in the fit and to a very
different definition of the NNLO charm contribution. 
There is agreement in the reduction of $\alpha_S(M_Z^2)$ at NNLO, i.e. 
$0.1171 \to 0.1143$. 

\vspace{-0.0cm}

\begin{figure}
\begin{center}
\includegraphics[width=.55\textwidth]{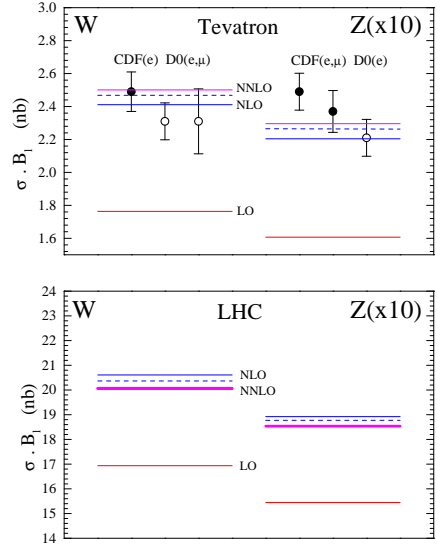}
\end{center}
\vspace{-1cm}
\caption[]{LO, NLO and NNLO predictions for $W$ and $Z$ cross-sections}
\label{eps29}
\end{figure}

\vspace{-0.0cm}

Using these NNLO partons there is reasonable stability order by order for 
the (quark-dominated) $W$ and $Z$ cross-sections, as seen in Fig.~8. 
However, the change from NLO to NNLO is of order $4\%$, which is much 
bigger than the uncertainty at NLO due to experimental errors. 
Also, this fairly good convergence is largely 
guaranteed because the quarks are fit directly to data.
There is greater danger in gluon dominated quantities, e.g. $F_L(x,Q^2)$, as 
shown in Fig.~9. Hence, the convergence from order to order is uncertain.

\begin{figure}
\begin{center}
\vspace{-0.5cm}
\includegraphics[width=.55\textwidth]{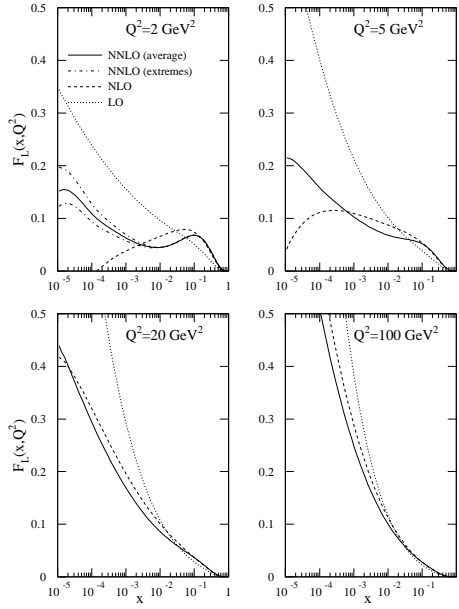}
\end{center}
\vspace{-0.6cm}
\caption[]{LO, NLO and NNLO predictions for $F_L(x,Q^2)$}
\label{eps30}
\end{figure}
\vspace{-0cm}

\vspace{-0.2cm}

\subsection{Empirical approach} 

We can estimate where theoretical errors may be important by adopting 
the empirical 
approach of investigating in detail the effect of cuts on the 
fit quality, i.e. we try varying the kinematic cuts on data. 
The procedure is to change $W^2_{cut}$, $Q^2_{cut}$ 
and/or $x_{cut}$, 
re-fit and see if the quality of the fit to the remaining data improves 
and/or the input parameters change dramatically. 
(This is similar to a previous suggestion in terms of 
data sets \cite{Collins}.) 
One then continues until the quality of the fit and the partons stabilize 
\cite{cuts}. 

\vspace{-0.0cm}

For $W^2_{cut}$ raising from $12.5 \GeV^2$ has no effect.
When raising $Q^2_{cut}$ from $2\GeV^2$ in steps there 
is a  slow, continuous and significant improvement for $Q^2$ up to 
$> 12 \GeV^2$ (631 data points cut) -- suggesting that 
any corrections are probably higher orders not higher twist. 
The input gluon becomes slightly smaller at low $x$
at each step (where one loses some of the lowest $x$ data), 
and larger at high $x$. 
$\alpha_S(M_Z^2)$ slowly decreases by about $0.0015$.
The fit improves for Tevatron jets and BCDMS data.   
Raising $x_{cut}$ leads to continuous improvement with  
stability reached at $x=0.005$ (271 data points cut) with 
$\alpha_S(M_Z^2) \to 0.118$. There is  
an improvement in the fit to HERA, NMC and Tevatron jet data, and 
much reduced tension between the data sets.
At each step the moderate $x$ gluon becomes more 
positive, at the expense of the gluon below the cut becoming very negative 
and $dF_2(x,Q^2)/d\ln Q^2$ being incorrect. However, higher orders could cure 
this in a quite plausible manner, e.g. adding higher order terms to 
the splitting functions 
$$ P_{gg} \to ....+ \frac{3.6\bar\alpha_S^4}{x}
\biggr(\frac{\ln^3(1/x)}{6}-
\frac{\ln^2(1/x)}{2}\biggr),$$
$$ P_{qg} \to ....+ \frac{4.3N_f\bar\alpha_S^5}{6x}
\biggr(\frac{\ln^3(1/x)}{6}-
\frac{\ln^2(1/x)}{2}\biggr),$$
leaves the fit above $x=0.005$ largely unchanged, but solves the problem 
below $x=0.005$. (Saturation corrections seem to make the fit worse.)   
Hence, the cuts are suggestive of theoretical errors for 
small $x$ and/or small $Q^2$. 
Predictions for $W$ and Higgs cross-sections at the Tevatron are still safe if 
$x_{cut}=0.005$, since they do not sample partons at lower $x$, 
and change in a smooth manner as $x_{cut}$ is lowered, due to the 
altered partons above $x_{cut}$, outside the 
limits set by experimental errors, as seen in Fig.~6.

\vspace{-0.2cm}

\section{Conclusions}

One can perform global fits to all up-to-date data over a wide range of 
parameter space, and  
there are various ways of looking at uncertainties
due to errors on data alone. There is no totally preferred approach. 
The errors from this source are rather small   
-- $\sim 1-5 \%$ except in a few regions of parameter space 
and are similar using various approaches. 
The uncertainty from input assumptions e.g. cuts on 
data, parameterizations {\it etc}., are comparable and sometimes 
larger, which means one cannot believe one group's errors.

The quality of the fit is fairly good, but there are some slight problems.
These imply that errors from higher orders/resummation are 
potentially large in some regions of parameter space,
and due to correlations between partons these affect all regions (the small 
$x$ gluon influences the large $x$ gluon). Cutting out low $x$ and/or $Q^2$ 
data allows a much-improved fit to the remaining data, and altered partons. 
Hence, for some processes theory is probably the dominant source of 
uncertainty at present and a systematic study is a priority.


\vspace{-0.2cm}

\begin{thebibliography}{xx}

\bibitem{MRST2001}A.D. Martin et~al., Eur. Phys. J. {\bf C23} 73 (2002).

\bibitem{CTEQ6} CTEQ Collaboration: J. Pumplin et~al., JHEP 0207:012 (2002).

\bibitem{Botje} M. Botje, Eur. Phys. J. {\bf C14} 285 (2000).

\bibitem{Giele} W.T. Giele and S. Keller, Phys. Rev. {\bf D58} 094023 (1998);
W.T. Giele, S. Keller and D.A. Kosower, {\tt hep-ph/0104052}.

\bibitem{Alekhin}S.I. Alekhin, Phys. Rev. {\bf D63} 094022 (2001).

\bibitem{H1Krakow} H1 Collaboration: C. Adloff et~al., Eur. Phys. J.
{\bf C21} 33 (2001).

\bibitem{ZEUSfit} A.M. Cooper-Sarkar, {\tt hep-ph/0205153}, J. Phys. 
{\bf G28} 2669 (2002); ZEUS Collaboration: S. Chekanov et~al., 
{\tt hep-ph/0208023}.

\bibitem{H1A} H1 Collaboration:  C. Adloff et~al., Eur. Phys. J.
{\bf C13} 609 (2000); H1 Collaboration:  C. Adloff et~al., Eur. Phys. J.
{\bf C19} 269 (2001).

\bibitem{ZEUS} ZEUS Collaboration:  S. Chekanov et~al., Eur.
Phys. J. {\bf C21} 443 (2001); ZEUS Collaboration: S. Chekanov et~al., 
{\tt hep-ph/0208040}.

\bibitem{E665} M.R. Adams et~al., Phys. Rev. {\bf D54} 3006 (1996).

\bibitem{BCDMS} BCDMS Collaboration:  A.C. Benvenuti et~al., Phys.
Lett. {\bf B223} 485 (1989); 
BCDMS Collaboration:  A.C. Benvenuti et~al., Phys.
Lett. {\bf B236} 592 (1989).

\bibitem{SLAC} L.W. Whitlow {\it et al}., Phys. Lett. {\bf B282} 475 (1992),
L.W. Whitlow, preprint SLAC-357 (1990).

\bibitem{NMC} NMC Collaboration:  M. Arneodo et~al., Nucl. Phys.
{\bf B483} 3 (1997); Nucl. Phys. {\bf B487} 3 (1997).

\bibitem{CCFR} CCFR Collaboration:  U.K. Yang et~al., Phys. Rev.
Lett. {\bf 86} 2742 (2001);
CCFR Collaboration:  W.G. Seligman  et~al., Phys. Rev.
Lett. {\bf 79} 1213 (1997).

\bibitem{ZEUSc} ZEUS Collaboration: J. Breitweg et~al.,
Eur. Phys. J. {\bf C12} 35 (2000).

\bibitem{H1c} H1 Collaboration:  C. Adloff et~al., Phys. Lett.
{\bf B528} 1999 (2002);

\bibitem{E605} E605 Collaboration: G. Moreno et~al., Phys. Rev.
{\bf D43} 2815 (1991).

\bibitem{E866} E866 Collaboration:  R.S. Towell et~al., Phys. Rev.
{\bf D64} 052002 (2001).

\bibitem{Wasymm} CDF Collaboration: F. Abe et~al., Phys. Rev. Lett.
{\bf 81} 5744 (1998).

\bibitem{D0} D0 Collaboration:  B. Abbott et~al., Phys. Rev. Lett.
{\bf 86} 1707 (2001).

\bibitem{CDF} CDF Collaboration:  T. Affolder et~al., Phys.
Rev. {\bf D64} 032001 (2001).

\bibitem{NuTeV} NuTeV Collaboration:  M. Goncharov et~al., {\tt
hep-ex/0102049}.

\bibitem{CTEQLag}CTEQ Collaboration: D. Stump et~al., Phys. Rev.
{\bf D65} 014012 (2002).

\bibitem{CTEQmul}CTEQ Collaboration:
J. Pumplin et~al., Phys. Rev. {\bf D65} 014011 (2002).

\bibitem{CTEQHes}CTEQ Collaboration:
J. Pumplin et~al., Phys. Rev. {\bf D65} 014013 (2002).

\bibitem{MRSTnew} A.D.Martin et~al {\tt hep-ph/0211080}.

\bibitem{THJPG}R.S. Thorne et~al., {\tt hep-ph/0205233},  J. Phys. 
{\bf G28} 2717 (2002).

\bibitem{Zomer}
  C. Pascaud and F. Zomer 1995 {\it Preprint} LAL-95-05.

\bibitem{NNLOmoms} S.A. Larin et~al., Nucl. Phys. {\bf B492} 338 (1997);
A. R\'{e}tey and J.A.M. Vermaseren, Nucl. Phys. {\bf
B604} 281 (2001).

\bibitem{NNLOsplit} W.L. van Neerven and A. Vogt, Phys. Lett. {\bf B490}
111 (2000).

\bibitem{MRSTNNLO} A.D. Martin et~al., Phys. Lett. {\bf B531} 216 (2002).

\bibitem{NNLOAl} S.I. Alekhin, Phys. Lett. {\bf B519} 57 (2001).

\bibitem{Collins} J. C. Collins and J. Pumplin, {\tt hep-ph/0105207}.

\bibitem{cuts} A.D. Martin, R.G. Roberts, W.J. Stirling and R.S.
Thorne, in preparation.

\end{thebibliography}
\end{document}